\documentclass[prd,onecolumn,nofootinbib,revtex4-2]{revtex4-2}

\setcitestyle{authoryear,round} 

\usepackage[utf8]{inputenc}
\usepackage[T1]{fontenc}
\usepackage{lmodern}
\usepackage{amsmath,amssymb}
\usepackage{graphicx}

\begin{document}
\title{The Roche-Lobe Radius in Newtonian and Non-Newtonian Gravity}

\author{Vel\'asquez-Toribio, A. M.}
\affiliation{Center for Astrophysics and Cosmology, Departamento de Física, Federal University of Esp\'irito Santo, 29075-910 Vit\'oria - ES, Brazil}
\email{alan.toribio@ufes.br}
\date{\today}

\begin{abstract}
In this paper, we extend the standard calculation of the Roche lobe from the Newtonian potential to non-Newtonian models, specifically Yukawa-type and power-law potentials. These models are significant because they arise from unification theories, such as those involving extra dimensions or scalar-tensor gravity. Furthermore, we develop fitting formulas for the equivalent volume radius ($R_L$) under these potentials, analogous to Eggleton’s formula for the Newtonian case. We calibrate the fitting formulas of the non-Newtonian potentials with their corresponding numerical solutions.
\end{abstract}

\maketitle

\section{Introduction}

The study of the Roche lobe is an essential tool for understanding the evolution and stability of binary systems, both stellar and planetary (Roche 1849; Kopal 1959; Paczynski 1971). In its classical formulation, the geometry of equipotential lobes and the location of the Lagrange points are derived under the assumption that gravity strictly follows the inverse-square law. Although this hypothesis has been verified with high precision within the Solar System, it is not an inevitable consequence of the general principles of gravitation and may be modified in different theoretical frameworks or at different distance scales.

Several contemporary unification schemes with extra-dimensional models, Randall--Sundrum-type gravity, and scalar--tensor extensions predict corrections to the Newtonian gravitational potential (Randall \& Sundrum 1999; Capozziello \& De Laurentis 2011). In many of these scenarios, the effective gravitational field acquires a modified behavior at intermediate or large distances, leading to interactions that can be slightly stronger or weaker than those prescribed by Newtonian gravity. Such deviations can be interpreted as macroscopic manifestations of additional degrees of freedom or of the curvature associated with extra dimensions, which alter the propagation of gravity beyond laboratory scales.

From a phenomenological perspective, the possibility of departures from the inverse-square law has been explored across a wide range of contexts, from torsion-balance experiments at sub-millimeter scales to astronomical distances (Iorio 2007; Lee et al.\ 2020; Adelberger et al.\ 2003, 2009; Fischbach \& Talmadge 1999; Sealfon et al.\ 2005; Klimchitskaya \& Mostepanenko 2012). In this setting, binary systems provide a privileged laboratory: the morphology of the Roche lobe and the stability of mass transfer depend directly on the functional form of the gravitational potential.

Analyzing the Roche lobe under non-Newtonian potentials therefore makes it possible to investigate, in a controlled manner, how small modifications to gravity can alter the topology of equipotential surfaces, shift equilibrium points, and modify the critical conditions for mass overflow. These studies are especially relevant in compact or exotic binaries, where the gravitational interaction may be subject to relativistic effects, scalar couplings, or long-range corrections (Soberman et al.\ 1997; Zhao \& Tian 2006; Sepinsky et al.\ 2007; Dai \& Blandford 2013; Frank et al.\ 2002; Dosopoulou \& Kalogera 2016; Zhao 2005; Morris 1994; De Laurentis et al.\ 2018; Idrisi et al.\ 2022).

In the Newtonian case, a key practical advance was Eggleton’s closed-form fit for the equivalent-volume Roche radius, which reproduces numerical equipotential calculations with high accuracy across a broad range of mass ratios (Eggleton 1983; Hilditch 2001). This compact expression has become standard in binary-evolution codes, enabling rapid evaluations inside iterative solvers and large population-synthesis pipelines without repeatedly solving for the full equipotential geometry.

Motivated by this success, it is natural to ask whether analogous fitting formulas can be constructed when gravity departs slightly from the inverse-square law. Developing Eggleton-type fits for Yukawa and power-law potentials would provide the same computational leverage in non-Newtonian settings: fast, accurate estimates of the equivalent-volume Roche radius that can be plugged directly into evolutionary calculations and parameter scans, while retaining clear control over the underlying physical modifications.

The Yukawa and power-law potentials considered below should be understood as phenomenological deformations of the point-mass Roche potential. They are not meant to include all physical corrections relevant in close interacting binaries. In particular, finite-size effects, tidal and rotational distortions, asynchronous rotation, eccentricity, and non-perturbative deformations of the stellar structure can produce corrections to the Roche geometry that are comparable to, or larger than, the modified-gravity contribution in sufficiently close systems. The purpose of the present work is more restricted: to determine how the critical equipotential surface and the volume-equivalent Roche-lobe radius respond when the central two-body potential itself is modified. In this sense, the results provide a controlled baseline for modified-potential effects, which may later be combined with more realistic close-binary models.

This article is organized as follows. In Sec.~II we review the derivation of the Roche lobe for Newtonian gravitation. Section~III develops the corresponding derivations for the Yukawa potential. Section~IV presents the case of a power-law potential. Section~V shows numerical solutions, and Sec.~VI summarizes our main conclusions.

\section{Newtonian gravity}

We begin by revisiting the classical construction of the Roche lobe in Newtonian gravity. The mathematical framework we use is the restricted circular three-body problem in a co-rotating frame, whose effective potential dictates the geometry of the equipotential surfaces. The Roche lobe is the critical, closed equipotential that touches the inner Lagrange point $L_1$. It is crucial to clarify the physical concept being calculated: we are not integrating the dynamic trajectory of a test particle. Instead, we are identifying this specific three-dimensional surface, which defines the gravitational limit that encloses the maximum volume that a quasi-hydrostatic fluid (the star) can occupy while remaining bound to a component.

We consider a circular binary with component masses $M_1$ and $M_2$ separated by a fixed distance $a$ (orbital separation). We work in a frame co-rotating with angular frequency
\begin{eqnarray}
\Omega^2 \;=\; \frac{G(M_1+M_2)}{a^3},
\end{eqnarray}
and frequently adopt dimensionless units $a=1$ and $G(M_1+M_2)=1$, so that $\Omega^2=1$. We also define the mass ratio and the reduced mass fraction as
\begin{eqnarray}
q \;\equiv\; \frac{M_1}{M_2},
\qquad
\mu \;\equiv\; \frac{M_2}{M_1+M_2} \;=\; \frac{1}{1+q}.
\end{eqnarray}

With this convention, no assumption is made about which component is more massive: if $q < 1$, the lobe associated with $M_1$ corresponds to the lower-mass component, whereas if $q > 1$, it corresponds to the higher-mass component.
In Cartesian coordinates $(x,y,z)$ with the origin at the system barycenter, the stars lie on the $x$ axis at
\begin{eqnarray}
x_1 \;=\; -\,\mu\,a,
\qquad
x_2 \;=\; (1-\mu)\,a.
\end{eqnarray}
The distances from a field point $\mathbf{r}=(x,y,z)$ to each star and the cylindrical radius in the co-rotating plane are
\begin{eqnarray}
r_1 \;=\; \sqrt{(x-x_1)^2+y^2+z^2},
\qquad
r_2 \;=\; \sqrt{(x-x_2)^2+y^2+z^2},
\qquad
R_\perp^2 \;=\; x^2+y^2.
\end{eqnarray}
The effective potential (gravity of $M_1$ and $M_2$ plus the centrifugal term) is
\begin{eqnarray}
\Phi_{\rm eff}(\mathbf{r})
\;=\;
-\,\frac{G M_1}{r_1}
\;-\;
\frac{G M_2}{r_2}
\;-\;
\frac{1}{2}\,\Omega^2\,R_\perp^2 .
\end{eqnarray}

On the $x$ axis ($y=z=0$), $L_1$ is determined by stationarity:
\begin{eqnarray}
\frac{\partial \Phi_{\rm eff}}{\partial x}(x,0,0) \;=\; 0,
\label{eq:L1_root_dimensional}
\end{eqnarray}
which becomes, in the usual normalized units ($a=1$, $G(M_1+M_2)=1$),
\begin{eqnarray}
\frac{M_1}{(x-x_1)^2}
\;-\;
\frac{M_2}{(x-x_2)^2}
\;-\; x
\;=\; 0,
\label{eq:L1_root}
\end{eqnarray}
the one-dimensional equation that yields the position of the Lagrange point $x_{L1}$.

Thus the Roche lobe around $M_1$ is the closed planar contour in $z=0$ satisfying
\begin{eqnarray}
g(x,y) \;=\; \Phi_{\rm eff}(x,y,0) - \Phi_{L1} \;=\; 0,
\label{eq:levelset}
\end{eqnarray}
where $\Phi_{L1}=\Phi_{\rm eff}(x_{L1},0,0)$. The desired Roche–lobe boundary in the plane is the level set $g(x,y)=0$. To determine this contour, we use the tangent-field method. By construction, the gradient $\nabla g=(\partial_x g,\partial_y g)$ is everywhere normal to this level set. A tangent vector is therefore obtained by a $90^\circ$ rotation of the gradient. One convenient choice is
\begin{eqnarray}
g(x,y) &\equiv& \Phi_{\rm eff}(x,y,0) - \Phi_{L1},
\\
\mathbf{t}(x,y)
&=&
\begin{pmatrix}\partial_y g\\[2pt]-\partial_x g\end{pmatrix}
\;=\;
\begin{pmatrix}\partial_y \Phi_{\rm eff}(x,y,0)\\[2pt]-\partial_x \Phi_{\rm eff}(x,y,0)\end{pmatrix}.
\end{eqnarray}
Let $(x(\ell),y(\ell))$ be a parametrization of the contour by an arclength–like parameter $\ell$. The tangent ordinary differential equation that marches along the boundary is
\begin{eqnarray}
\frac{d}{d\ell}\begin{pmatrix}x\\[2pt] y\end{pmatrix}
&=&
\begin{pmatrix}\partial_y \Phi_{\rm eff}(x,y,0)\\[2pt]-\partial_x \Phi_{\rm eff}(x,y,0)\end{pmatrix},
\qquad
g\!\bigl(x(0),y(0)\bigr)=0.
\label{eq:tangent_ode}
\end{eqnarray}
The initial point $(x(0),y(0))$ must lie on the level set $g=0$ (for example, on the $x$ axis just inside the lobe).

Along any smooth solution of \eqref{eq:tangent_ode}, the chain rule gives
\begin{eqnarray}
\frac{d}{d\ell}\,g\!\bigl(x(\ell),y(\ell)\bigr)
&=&
\nabla g\cdot \frac{d}{d\ell}\begin{pmatrix}x\\ y\end{pmatrix}
=
\begin{pmatrix}\partial_x g\\ \partial_y g\end{pmatrix}\!\cdot\!
\begin{pmatrix}\partial_y g\\ -\partial_x g\end{pmatrix}
=
\partial_x g\,\partial_y g-\partial_y g\,\partial_x g
\;=\; 0,
\end{eqnarray}
hence $g$ remains constant along the trajectory; if $g=0$ at $\ell=0$, then $g=0$ for all $\ell$, and the curve traced by \eqref{eq:tangent_ode} stays on the desired level set.

If one wishes to use true arclength $k$ as parameter, the tangent field can be normalized:
\begin{eqnarray}
\frac{d}{dk}\begin{pmatrix}x\\ y\end{pmatrix}
&=&
\frac{1}{\|\nabla\Phi_{\rm eff}(x,y,0)\|}
\begin{pmatrix}\partial_y \Phi_{\rm eff}(x,y,0)\\[2pt]-\partial_x \Phi_{\rm eff}(x,y,0)\end{pmatrix},
\label{contorno}
\end{eqnarray}
so that $\bigl\|\tfrac{d}{dk}(x,y)\bigr\|=1$. This reparametrization does not change the path, only the speed at which it is traversed. A single Newton projection step along the normal,
\begin{eqnarray}
\begin{pmatrix}x\\ y\end{pmatrix}
&\leftarrow&
\begin{pmatrix}x\\ y\end{pmatrix}
\;-\;
\frac{g(x,y)}{\|\nabla g(x,y)\|^2}\,\nabla g(x,y).
\label{contornocorrection}
\end{eqnarray}

In the numerical implementation, we integrate the normalized tangent equation using a fourth-order Runge--Kutta scheme in dimensionless units ($a = 1$ and $G(M_1 + M_2) = 1$). The arclength-like integration step is set to a fixed value $\Delta k$. A Newton projection step, following Eq. (14), is applied periodically and also whenever the level-set residual exceeds a predefined tolerance $|g|$. The contour is regarded as closed when the integration returns to the neighborhood of the initial point with a matching tangent direction. For the volume-equivalent radius, the upper branch $y_{\max}(x)$ is reconstructed by bisection on the level-set equation over a uniform grid in $x$. Sensitivity tests performed by refining the integration step and the grid density indicate that the equivalent-volume radius $R_L/a$ remains stable within the required numerical precision for the cases shown.

It is useful to verify the asymptotic value of $r$ for $q\ll 1$, because it fixes the correct behavior in a controlled regime and allows us to benchmark against the numerical results. Returning to the equation for the inner Lagrange point, seek $L_1$ near $M_1$ by writing $x=-\mu+r$ with $0<r\ll 1$. Then
\begin{eqnarray}
x-x_1 \;=\; r,
\qquad
x-x_2 \;=\; -(1-r),
\qquad
\frac{M_1}{(x-x_1)^2} \;=\; \frac{M_1}{r^2},
\qquad
\frac{M_2}{(x-x_2)^2} \;=\; \frac{M_2}{(1-r)^2}.
\end{eqnarray}
Thus, the $L_1$ equation \eqref{eq:L1_root} becomes
\begin{eqnarray}
\frac{q}{r^2}
\;-\;
\frac{1}{(1-r)^2}
\;+\;
1
\;-\;
(1+q)\,r
\;=\;
0.
\label{eq:L1_aux}
\end{eqnarray}
Expanding $(1-r)^{-2}$ and keeping dominant terms for $r\ll1$ gives
\begin{eqnarray}
r^3 \;\simeq\; \frac{q}{3}.
\end{eqnarray}

\subsection{The volume-equivalent Roche radius}

The Roche lobe around a star in a circular binary system is generally not spherical. Therefore, to usefully summarize its size, we often employ the volume-equivalent radius: the radius of a sphere that would have the same volume as the lobe. This is the quantity typically used in many population and stability estimates because it compares directly to the stellar radius.

Let $\mathcal{S}$ denote the closed level set (equipotential) that goes through $L_1$:
\begin{eqnarray}
\mathcal{S} \;=\; \Big\{ (x,y,z)\ :\ \Phi_{\rm eff}(x,y,z) \;=\; \Phi_{\rm eff}(L_1) \Big\}.
\end{eqnarray}
The \emph{Roche-lobe volume} $V$ is the volume enclosed by $\mathcal{S}$, and the volume-equivalent radius is defined by
\begin{eqnarray}
R_L \;\equiv\; \left(\frac{3V}{4\pi}\right)^{1/3}.
\label{eq:def_RL}
\end{eqnarray}
The Roche–lobe contour in the orbital plane ($z=0$) is the level set of the effective potential at the $L_1$ value, i.e.\ $\{(x,y):\, \Phi_{\rm eff}(x,y,0)=\Phi_{L1}\}$. For each $x$ where the closed contour exists, it intersects the vertical line at two symmetric points $(x,\pm y_{\max}(x))$. We define the upper envelope $y_{\max}(x)\ge 0$ as the positive branch (the portion with $y\ge 0$), over the interval $x\in[x_{\min},x_{\max}]$ that spans the lobe. Consider a thin vertical strip at position $x$, of thickness $dx$. Rotating this strip about the $x$ axis generates a thin disk of radius $y_{\max}(x)$ and thickness $dx$. The disk’s volume is
\begin{eqnarray}
dV \;=\; \pi\,\big[y_{\max}(x)\big]^2\,dx.
\end{eqnarray}
The integration over the range $x_{\min}$ to $x_{\max}$ provides the complete three-dimensional volume of the lobe:
\begin{eqnarray}
V \;=\; \pi \int_{x_{\min}}^{x_{\max}} y_{\max}(x)^2\,dx.
\label{eq:V_washers}
\end{eqnarray}

When the mass ratio $q=M_1/M_2$ is small and the lobe is taken around $M_1$, the $L_1$ point lies a short distance $r$ from $M_1$. A local analysis shows that, in coordinates rescaled by $r$, the dimensionless lobe shape tends to a universal profile. Consequently, the enclosed volume scales as
\begin{eqnarray}
V \;=\; C\,r^3,
\label{eq:V_Cr3}
\end{eqnarray}
where $C$ is a pure constant fixed by the universal shape. Combining \eqref{eq:V_Cr3} with \eqref{eq:def_RL} gives
\begin{eqnarray}
\frac{R_L}{a}
\;=\;
\left(\frac{3C}{4\pi}\right)^{1/3}\frac{r}{a}.
\label{eq:RL_from_r}
\end{eqnarray}
Using the result for $q \ll 1$ we write
\begin{eqnarray}
\frac{R_L}{a}
\;=\;
\chi
\left(\frac{q}{3}\right)^{1/3},
\label{eq:RL_smallq}
\end{eqnarray}
where the factor $\chi=\left(\tfrac{3C}{4\pi}\right)^{1/3}$ is a weakly model-dependent geometric constant that converts the local throat scale $r$ into the global volume. In Newtonian gravity, a direct three-dimensional integration gives $\chi\simeq 0.667$, i.e.\ $R_L/a\simeq 0.462\,q^{1/3}$ for $q\ll 1$.

\subsection{The Eggleton approximation}

Eggleton's closed form for the lobe around $M_1$ is (Eggleton 1983)
\begin{eqnarray}
\frac{R_L}{a} \;\approx\;
\frac{0.49\,q^{2/3}}{0.6\,q^{2/3}+\ln\!\bigl(1+q^{1/3}\bigr)}.
\label{eggleton}
\end{eqnarray}
For $q\to0$, using $\ln(1+q^{1/3})=q^{1/3}-\tfrac{1}{2}q^{2/3}+O(q)$,
\begin{eqnarray}
\frac{R_L}{a}
\;=\;
\frac{0.49\,q^{2/3}}{q^{1/3}+0.1\,q^{2/3}+O(q)}
\;=\;
0.49\,q^{1/3}\,\bigl[1+O(q^{1/3})\bigr],
\end{eqnarray}
i.e., it reproduces the same $q^{1/3}$ scaling. Identifying $0.49=\chi\,3^{-1/3}$ yields $\chi\simeq 0.49\cdot 3^{1/3}\simeq 0.707$, the effective value associated with Eggleton's global fit (thus slightly larger than, $\chi\simeq0.667$).

Let $E(q)$ denote the right-hand side of \eqref{eggleton}. The lobe radius around $M_1$ is $R_{L,1}/a=E(q)$, while the lobe around $M_2$ corresponds to inverting the mass ratio:
\begin{eqnarray}
\frac{R_{L,2}}{a} \;=\; E\!\left(\frac{1}{q}\right).
\end{eqnarray}
Thus, the transformation $q\leftrightarrow 1/q$ does not leave $R_{L,1}$ invariant, but maps the lobe of one component to that of the other, as required by the physical symmetry of the binary. This provides a practical consistency check and ties together the extreme-mass-ratio limits. At $q=1$, the symmetry enforces $R_{L,1}=R_{L,2}$.

\section{The Yukawa potential}

A practical and widely used phenomenological way to test departures from Newtonian gravity is the Yukawa potential, which encapsulates short-range deviations driven by massive force carriers, extra dimensions, or light scalar fields with screening. Even modest values of the strength \(\alpha\) and range \(\lambda\) can reshape the near-field balance that locates \(L_1\), thereby altering the Roche-lobe size and the onset and stability of mass transfer, key ingredients for population-synthesis studies of interacting binaries. For background and constraints on Yukawa-like forces see, e.g., Adelberger et al.\ (2003, 2009) and Fischbach \& Talmadge (1999). The form of the potential and the force is given by:
\begin{eqnarray}
\Phi(r) \;=\; -\,\frac{G M}{r}\,\Big[\,1+\alpha\,e^{-\,r/\lambda}\,\Big],
\qquad
F(r) \;=\; -\,\frac{G M}{r^2}\,\Big[\,1+\alpha\Big(1+\frac{r}{\lambda}\Big)e^{-\,r/\lambda}\,\Big].
\end{eqnarray}
The parameter $\alpha$ is the coupling strength of the Yukawa correction (attractive if $\alpha>0$, repulsive if $\alpha<0$). The parameter $\lambda$ is the interaction range. For distances $r\ll \lambda$ the correction is strong, whereas for $r\gg\lambda$ the force is effectively Newtonian. We adopt dimensionless variables
\begin{eqnarray}
\Lambda \;\equiv\; \frac{\lambda}{a},
\qquad
s \;\equiv\; \frac{r}{a},
\end{eqnarray}
and define
\begin{eqnarray}
\Upsilon(s;\alpha,\Lambda) \;\equiv\; 1+\alpha\Big(1+\frac{s}{\Lambda}\Big)e^{-\,s/\Lambda},
\qquad
\Upsilon_0 \equiv \Upsilon(0)=1+\alpha,
\qquad
\Upsilon_1 \equiv \Upsilon(1;\alpha,\Lambda).
\end{eqnarray}
For a circular orbit with separation $a$ one finds
\begin{eqnarray}
\Omega^2 \;=\; \frac{G(M_1+M_2)}{a^3}\,\Upsilon_1.
\end{eqnarray}
In the normalized units used earlier ($a=1$, $G(M_1+M_2)=1$), this reduces to
\begin{eqnarray}
\Omega^2 \;=\; \Upsilon_1.
\end{eqnarray}

In the co-rotating frame (restricted to the orbital plane $z=0$), the effective potential is
\begin{eqnarray}
\Phi_{\rm eff}(x,y,0)
&=&
-\,\frac{M_1}{s_1}\Big[1+\alpha\,e^{-\,s_1/\Lambda}\Big]
-\,\frac{M_2}{s_2}\Big[1+\alpha\,e^{-\,s_2/\Lambda}\Big]
-\,\frac{1}{2}\,\Omega^2\big(x^2+y^2\big),
\end{eqnarray}
with in-plane distances
\begin{eqnarray}
s_1 &\equiv& \sqrt{(x-x_1)^2+y^2},
\qquad
s_2 \;\equiv\; \sqrt{(x-x_2)^2+y^2}.
\end{eqnarray}
The in-plane derivatives take the compact form
\begin{eqnarray}
\frac{\partial\Phi_{\rm eff}}{\partial x}(x,y,0)
&=&
M_1\,\frac{x-x_1}{s_1^{3}}\,\Upsilon(s_1)
\;+\;
M_2\,\frac{x-x_2}{s_2^{3}}\,\Upsilon(s_2)
\;-\;\Omega^2\,x,
\\
\frac{\partial\Phi_{\rm eff}}{\partial y}(x,y,0)
&=&
M_1\,\frac{y}{s_1^{3}}\,\Upsilon(s_1)
\;+\;
M_2\,\frac{y}{s_2^{3}}\,\Upsilon(s_2)
\;-\;\Omega^2\,y.
\end{eqnarray}
As in the Newtonian case, the lobe contour in the orbital plane is the level set
\begin{eqnarray}
g(x,y) &\equiv& \Phi_{\rm eff}(x,y,0)-\Phi_{L1} \;=\; 0,
\qquad
\Phi_{L1} \;\equiv\; \Phi_{\rm eff}(x_{L1},0,0).
\label{yukawacondition}
\end{eqnarray}
The method to determine the two-dimensional contours and the three-dimensional volume is the same as before in the Newtonian case; only the effective potential is replaced by the Yukawa form. Consequently, we employ Eqs.~\eqref{contorno} and \eqref{contornocorrection} together with \eqref{yukawacondition}.

It is also instructive to verify the asymptotic behavior in the Yukawa case. Place $M_1$ at $x_1=-\mu$ and $M_2$ at $x_2=1-\mu$ with $\mu=1/(1+q)$, and seek $L_1$ as $x=-\mu+s$ with $0<s\ll1$. On the $x$ axis the stationarity reads
\begin{eqnarray}
\frac{\partial\Phi_{\rm eff}}{\partial x}=0
&\Longrightarrow&
\frac{(1-\mu)\,\Upsilon(s)}{s^{2}}
\;-\;
\frac{\mu\,\Upsilon(1-s)}{(1-s)^{2}}
\;-\;
\Omega^2\,(-\mu+s)
\;=\; 0.
\label{eq:Yukawa_L1_full}
\end{eqnarray}
Expanding near $s=0$,
\begin{eqnarray}
\Upsilon(s) &\simeq& \Upsilon_0,
\qquad
\Upsilon(1-s) \;\simeq\; \Upsilon_1 - \Upsilon'_1\,s,
\qquad
\Upsilon'_1 \;\equiv\; \left.\frac{d\Upsilon}{ds}\right|_{s=1}
\;=\; -\,\frac{\alpha\,e^{-1/\Lambda}}{\Lambda^{2}},
\end{eqnarray}
and using $\Omega^2=\Upsilon_1$, one obtains (after the constant pieces cancel)
\begin{eqnarray}
s^3 &\simeq& \frac{q\,\Upsilon_0}{\big(3+\zeta\big)\,\Upsilon_1},
\qquad
\zeta \;\equiv\; \frac{\Upsilon'_1}{\Upsilon_1}
\;=\;
-\,\frac{\alpha\,e^{-1/\Lambda}}{\Lambda^{2}\,\Upsilon_1}.
\label{eq:Yukawa_r_cubic}
\end{eqnarray}
Equation~(\ref{eq:Yukawa_r_cubic}) shows that the Newtonian $q/3$ result is multiplied by the near-body factor $\Upsilon_0$, by the inter-body factor $\Upsilon_1$ (entering $\Omega$), and by a small gradient correction $\zeta$.

It is important to specify the length scale probed by the Roche surface. Thus, considering the dimensionless variables defined at the beginning of this section, the Yukawa correction depends on 
$s/\Lambda=r/\lambda$. The fractional correction to the Newtonian force is:

\begin{eqnarray}
\epsilon_Y(s)=|\alpha|\left(1+\frac{s}{\Lambda}\right)e^{-s/\Lambda}.
\end{eqnarray}
Thus, the correction is relevant on the Roche surface only if the typical lobe scale, $s\sim R_L/a$, is comparable to or smaller than $\Lambda$. For $s\gg\Lambda$, the model rapidly approaches Newtonian gravity, whereas for $\Lambda\gg 1$, the correction is nearly constant across the binary and can be partially degenerate with a rescaling of the gravitational strength.

\subsection{Eggleton-like fitting formula}

It is useful to derive a Yukawa analogue of Eggleton’s fitting formula. In the small-lobe limit near $M_1$, one separates a ``local'' factor (the field of $M_1$) from the ``orbital'' factors (the orbital frequency and the tide due to $M_2$ at $r=a$). Expanding the effective potential around $M_1$ up to quadratic order and proceeding as in the Newtonian case identifies the cubic scale
\begin{eqnarray}
r^3 \;=\; a^3\,
\frac{q\,\Upsilon_0}{(3+\zeta)\,\Upsilon_1},
\end{eqnarray}
so that
\begin{eqnarray}
\frac{R_L}{a} \;\simeq\; \chi_Y(\alpha,\Lambda)\,
\left[\frac{q\,\Upsilon_0}{(3+\zeta)\,\Upsilon_1}\right]^{1/3}
\qquad (q\ll 1).
\end{eqnarray}
To extend to arbitrary $q$ while remaining compatible with numerics, we adopt the $q$-power multiplicative ansatz on top of the Eggleton baseline:
\begin{eqnarray}
\frac{R_L}{a}(q;\lambda,\alpha)
&=& \frac{R_L^{\mathrm{Egg}}(q)}{a}\,
\Bigl[\,1 + d_1(\lambda,\alpha)\,q^{1/3}
        + d_2(\lambda,\alpha)\,q^{2/3}
        + d_3(\lambda,\alpha)\,q
        + \mathcal{O}\!\bigl(q^{4/3}\bigr)\Bigr],
\label{eq:fit_mq}
\end{eqnarray}
where $R_L^{\mathrm{Egg}}/a$ is given by \eqref{eggleton} and the coefficients $d_i(\lambda,\alpha)$ encode the Yukawa deviation. This preserves the leading $q^{1/3}$ exponent and smoothly modulates the prefactor. Under $q\mapsto 1/q$, the companion’s lobe is obtained by applying the same multiplicative factor to $E(1/q)$, so the exchange symmetry is respected.

The coefficients $d_i(\lambda,\alpha)$ are obtained by fitting Eq.~\eqref{eq:fit_mq} to the numerically computed Roche-lobe radii for a grid of Yukawa parameters $(\alpha,\lambda)$. For each pair $(\alpha,\lambda)$, we calculate the equipotential contour at $\Phi_{L1}$ and derive the equivalent-volume radius $R_L/a$. The residuals with respect to the Eggleton baseline are then expanded in powers of $q^{1/3}$ and minimized in a least-squares sense to determine $d_1$, $d_2$, and $d_3$. This procedure captures both the strength ($\alpha$) and range ($\lambda$) dependencies smoothly, enabling interpolation across the parameter space for binaries.

\section{The power-law potential}

Another non-Newtonian potential of importance is the power-law potential, often inspired by theories that test extensions of fundamental physics, such as models involving large extra dimensions. In this scenario, one can either completely replace the Newtonian potential with a power law proportional to $r^{-(2+\nu)}$, or one can consider perturbations over the Newtonian potential. In the present article, we choose to consider the case of perturbations over the Newtonian potential, primarily motivated by the fact that theories like Randall--Sundrum specifically predict the Newtonian potential plus power-law perturbations. Thus we model a small deviation from Newton as
\begin{eqnarray}
\Phi(r) &=& -\,\frac{G M}{r}\,\Big[\,1 + \beta\,\big(\tfrac{r}{a}\big)^{-p}\,\Big],
\qquad 0<|\beta|\ll 1,\ \ p>0,
\label{eq:Phi_def}
\end{eqnarray}
where $a$ is the binary separation used as reference scale, $p$ is the index of the correction, and $\beta$ its (dimensionless) amplitude. The central force is then
\begin{eqnarray}
\vec{F}(r)
&=&
-\,\frac{G M}{r^{2}}\,
\Big[\,1 + \beta\,(1{+}p)\,\big(\tfrac{a}{r}\big)^{p}\,\Big]\;\hat{r}.
\label{eq:F_central}
\end{eqnarray}
Introducing $s\equiv r/a$ and the (dimensionless) force modifier
\begin{eqnarray}
\Upsilon(s) &\equiv& 1 + \beta\,(1{+}p)\,s^{-p},
\label{eq:Upsilon_def}
\end{eqnarray}
the Newtonian $1/r^2$ law is multiplied by $\Upsilon(s)$ and the circular frequency becomes
\begin{eqnarray}
\Omega^2 &=& \frac{G(M_1{+}M_2)}{a^{3}}\;\Upsilon(1)
\;=\;
\frac{G(M_1{+}M_2)}{a^{3}}\;\Big[\,1+\beta(1{+}p)\,\Big].
\end{eqnarray}
In normalized units ($a=1$, $G(M_1+M_2)=1$) this reduces to
\begin{eqnarray}
\Omega^2 &=& \Upsilon(1) \;=\; 1+\beta(1{+}p).
\label{eq:Omega2}
\end{eqnarray}

Place $M_1$ at $x_1=-\mu$ and $M_2$ at $x_2=1-\mu$, with $q=M_1/M_2$ and $\mu=1/(1+q)$. On the orbital plane $z=0$, with $s_1=\sqrt{(x-x_1)^2+y^2}$ and $s_2=\sqrt{(x-x_2)^2+y^2}$, the effective potential is
\begin{eqnarray}
\Phi_{\rm eff}(x,y,0)
&=&
-\,\frac{M_1}{s_1}\,\Big[1 + \beta\,s_1^{-p}\Big]
-\,\frac{M_2}{s_2}\,\Big[1 + \beta\,s_2^{-p}\Big]
-\,\frac{1}{2}\,\Omega^2\,(x^2+y^2).
\label{eq:Phi_eff}
\end{eqnarray}
Using \eqref{eq:Upsilon_def}, the in-plane derivatives take the compact form
\begin{eqnarray}
\frac{\partial\Phi_{\rm eff}}{\partial x}
&=&
M_1\,\frac{x-x_1}{s_1^{3}}\,\Upsilon(s_1)
\;+\;
M_2\,\frac{x-x_2}{s_2^{3}}\,\Upsilon(s_2)
\;-\;\Omega^2\,x,
\\
\frac{\partial\Phi_{\rm eff}}{\partial y}
&=&
M_1\,\frac{y}{s_1^{3}}\,\Upsilon(s_1)
\;+\;
M_2\,\frac{y}{s_2^{3}}\,\Upsilon(s_2)
\;-\;\Omega^2\,y,
\end{eqnarray}
which mirrors the Newtonian expressions with the multiplicative factor $\Upsilon$.

On the $x$ axis ($y=0$) between the masses ($x_1<x<x_2$), the stationarity condition $\partial_x\Phi_{\rm eff}=0$ yields the exact one-dimensional equation for $L_1$:
\begin{eqnarray}
M_1\,\frac{x-x_1}{|x-x_1|^{3}}\,\Upsilon(|x-x_1|)
\;+\;
M_2\,\frac{x-x_2}{|x-x_2|^{3}}\,\Upsilon(|x-x_2|)
\;-\;\Omega^2\,x
\;=\; 0.
\label{eq:L1_exact_pert}
\end{eqnarray}
In the case $q\ll 1$, let $x=-\mu+r$ with $0<r\ll 1$. Then $|x-x_1|=r$ and $|x-x_2|=1-r$. Expanding to first order in $r$ and to first order in $\beta$ one obtains
\begin{eqnarray}
\frac{q}{(1+q)}\,\frac{\Upsilon(r)}{r^{2}}
&\simeq&
\big(3 + \zeta\big)\,\Omega^2\,r,
\qquad
\zeta \;\equiv\; \frac{\Upsilon'(1)}{\Upsilon(1)} \;=\; -\,\beta\,p(1{+}p)\;+\;\mathcal{O}(\beta^2),
\label{eq:balance_smallq}
\end{eqnarray}
where we used $\Upsilon(r)=1+\beta(1{+}p)\,r^{-p}$ and \eqref{eq:Omega2}. Keeping terms up to $\mathcal{O}(\beta)$ and using $q/(1+q)\simeq q$, it is convenient to reference the Newtonian scale $r_N$ defined by
\begin{eqnarray}
r_N^3 &=& \frac{q}{3},
\end{eqnarray}
and write the first-order correction as
\begin{eqnarray}
r &=& r_N\;\Big[\,1 \;-\; \frac{\beta(1{+}p)}{3}\,\big(1 - r_N^{-p}\big)\,\Big]\;+\;\mathcal{O}(\beta^2).
\label{eq:r_first_order_beta}
\end{eqnarray}

A quadratic expansion of $\Phi_{\rm eff}$ about $L_1$ shows that the characteristic linear dimensions of the equipotential through $L_1$ scale $\propto r$, hence the enclosed volume scales as $V=C\,r^3$ with a dimensionless $C$ that varies smoothly with $(p,\beta)$ and tends to the Newtonian value as $\beta\to 0$. Defining $R_L$ by $V=(4\pi/3)R_L^3$ gives
\begin{eqnarray}
R_L &=& \chi\,r,
\qquad
\chi \;=\; \Big(\tfrac{3C}{4\pi}\Big)^{1/3},
\end{eqnarray}
so that, for $q\ll 1$,
\begin{eqnarray}
\frac{R_L}{a}
&\simeq&
\chi(0)\,r_N\;\Big[\,1 \;-\; \frac{\beta(1{+}p)}{3}\,\big(1 - r_N^{-p}\big)\,\Big]
\;+\;\mathcal{O}(\beta^2),
\qquad
r_N=\Big(\frac{q}{3}\Big)^{1/3},
\label{eq:RL_smallq_beta}
\end{eqnarray}
with $\chi(0)\simeq 0.667$ in the Newtonian limit. The factor $\chi$ inherits $\mathcal{O}(\beta)$ corrections but remains smooth for small $\beta$.

For arbitrary $q$, solve Eq.~\eqref{eq:L1_exact_pert} for $x_{L1}$, evaluate $\Phi_{L1}=\Phi_{\rm eff}(x_{L1},0,0)$, and extract the 2D/3D lobe from the level set $\Phi_{\rm eff}=\Phi_{L1}$ using the same method as in the Newtonian and Yukawa cases. The enclosed volume $V(q,p,\beta)$ then defines
\begin{eqnarray}
R_L(q,p,\beta) &=& \left(\frac{3\,V(q,p,\beta)}{4\pi}\right)^{1/3}.
\end{eqnarray}

For the power-law model the fractional force correction is
\begin{eqnarray}
\epsilon_P (s) & = & |\beta|(1 + p)s^{-p}.
\end{eqnarray}
Consequently, the perturbative interpretation requires $|\beta|(1 + p)s_{min}^{-p} \ll 1$ over the 
portion of the Roche surface being used. If this condition is not satisfied, the power-
law potential should be interpreted only as an effective parametrization of a stronger 
short-distance deformation, rather than as a small perturbation of Newtonian gravity.

\subsection{Eggleton-like fitting formula}

We consider a small power-law correction to Newtonian gravity, parameterized by an amplitude $\beta$ and index $p>0$. In this perturbative regime a convenient, accurate form is to keep the Eggleton baseline and multiply it by a smooth series in fractional powers of $q$:
\begin{eqnarray}
\frac{R_L}{a}(q;\,p,\beta)
&=&
\frac{R_L^{\mathrm{Egg}}(q)}{a}\;
\Bigl[
1 + d_1(p,\beta)\,q^{1/3}
  + d_2(p,\beta)\,q^{2/3}
  + d_3(p,\beta)\,q
  + \mathcal{O}\!\bigl(q^{4/3}\bigr)
\Bigr].
\label{eq:PL_ansatz}
\end{eqnarray}
Here $R_L^{\mathrm{Egg}}/a$ is the Newtonian Eggleton radius, and the coefficients $d_i(p,\beta)$ encode the power-law deviation. They are smooth functions of $(p,\beta)$ and satisfy
\begin{eqnarray}
d_i(p,0) \;=\; 0
\qquad
\text{for all }i,
\label{eq:PL_limit_Newton}
\end{eqnarray}
so that the model reduces continuously to Eggleton when $\beta\to 0$.

For small $q$,
\begin{eqnarray}
\frac{R_L^{\mathrm{Egg}}}{a} \;\sim\; 0.49\,q^{1/3}
\qquad (q\to 0),
\end{eqnarray}
hence
\begin{eqnarray}
\frac{R_L}{a}(q;\,p,\beta)
\;\sim\;
0.49\,q^{1/3}\,\Bigl[\,1+\mathcal{O}(q^{1/3})\Bigr]
\qquad (q\to 0),
\label{eq:PL_small_q}
\end{eqnarray}
which preserves the physically required exponent $1/3$ in the perturbative power-law setting.

The calibration of the coefficients $d_i(p,\beta)$ is straightforward at fixed $(p,\beta)$: given numerical values $\{(q_j,R_L^{\mathrm{num}}(q_j;p,\beta))\}_{j=1}^N$, one fits the multiplicative factor in \eqref{eq:PL_ansatz} by least squares on $q^{1/3}$, $q^{2/3}$, and $q$, reusing the same procedure adopted for the Yukawa case. Under $q\leftrightarrow 1/q$, the companion’s lobe is obtained analogously by evaluating the same multiplicative factor at $1/q$, just as in the Yukawa discussion above.

\section{Results and Discussion}

Figure 1 shows the two-dimensional Roche-lobe contours obtained for the three gravitational models considered in this work: the Newtonian case (solid curves), the Yukawa potential (dashed curves), and the power-law correction (dotted curves). Each panel presents the equipotential surface through the $L_1$ point in the orbital plane for four mass ratios: $q = 0.0400$, $0.5000$, $1.0000$, and $1.2000$. The dimensionless distances $s = r/a$ probed by these surfaces are of order $R_L/a$, which, according to the Newtonian Eggleton estimate, ranges from $\simeq 0.159$ for $q = 0.0400$ to $0.395$ for $q = 1.2000$. Consequently, these examples primarily test the modified potentials in the range $s \sim 0.1$--$0.4$, where the Yukawa correction is relevant only if the scale $\Lambda = \lambda/a$ overlaps the Roche scale, and the power-law correction remains perturbative only if $|\beta|(1 + p)s^{-p} \ll 1$ on the relevant portion of the lobe. 

For the extreme mass-ratio case ($q = 0.0400$), the non-Newtonian models produce small, yet visible, shifts in the size and shape of both lobes, while remaining in close proximity to the Newtonian contour. As the mass ratio $q$ approaches unity, the three potentials yield almost indistinguishable Roche lobes on both sides of the binary. This outcome illustrates that, for this particular selection of Yukawa and power-law parameters, the non-Newtonian corrections can be tuned to be competitive with the Newtonian prediction across a broad spectrum of mass ratios, with the full two-lobe geometry involving distances only up to order unity.

Figure 2 shows the full three-dimensional reconstruction of the Newtonian Roche
lobes for two representative mass ratios, q = 0.0400 and q = 1.2000.
In each panel, the closed equipotential surface through $L_1$,
computed in the orbital plane and then revolved around the $x$ axis as
described in the volume–equivalent construction, defines the Roche lobe around
each component. For the extreme mass–ratio case ($q=0.0400$), the lobe around
the low–mass is much smaller and more strongly curved than the extended
lobe of the massive companion, illustrating the $R_L/a \propto q^{1/3}$ scaling
derived analytically. For $q=1.200$, the two lobes become comparable in size
and nearly mirror–symmetric about the system barycenter, as expected for nearly
equal–mass binaries. 

Figure 3 displays the three-dimensional Roche lobes for the Yukawa 
potential, using the same representative mass ratios and the parameters   
$\Lambda = 0.3$ and $\alpha = 1.5$. The overall geometry of the lobes
remains very similar to the Newtonian reference, but both the size and the
curvature of the surfaces are now controlled by the effective factors
$\Upsilon_0$ and $\Upsilon_1$ discussed in Sec. III. For the
extreme mass–ratio example ($q=0.0440$), the low–mass lobe is slightly
inflated compared to the Newtonian one, reflecting the stronger effective
gravity in the near field of $M_1$, while the extended lobe around the massive
companion adjusts so as to preserve the $L_1$ balance. For $q=1.200$, the two
Yukawa lobes remain nearly symmetric about the barycenter, but their volumes
deviate at the percent level from the Newtonian values, illustrating how a
fixed pair $(\alpha,\lambda)$ can produce almost degenerate projected contours
while still inducing measurable changes in the equivalent-volume radius.

Figure 4 illustrates the three-dimensional Roche lobes obtained 
for the power-law correction to Newtonian gravity, using $p = 1.0$ and $\beta = 0.3$.
 As in the Yukawa case, the overall morphology of the
lobes closely resembles the Newtonian reference, but their volumes are
systematically shifted by the multiplicative factor $\Upsilon(s)$ that enters
the effective force.  These examples complement the
Yukawa reconstructions by showing that the same numerical machinery can track
smooth, model–dependent deformations of the full 3D Roche geometry across
different classes of non–Newtonian potentials.

Since the three-dimensional Roche-lobe surfaces remain visually close to
their Newtonian counterparts, we complement the geometric comparison with a
local residual diagnostic in the orbital plane. For each component
$M_i$, located at the barycentric position ${\bf x}_i=(x_i,0)$, we
introduce a polar parametrization centered on that component. The angular
origin is chosen along the direction from $M_i$ to the inner Lagrange point
$L_1$. We therefore define
\begin{eqnarray}
{\bf e}_{L_1}^{(i)}
&=&
\frac{{\bf x}_{L_1}-{\bf x}_i}
{|{\bf x}_{L_1}-{\bf x}_i|},
\label{eq:eL1_direction}
\end{eqnarray}
where ${\bf e}_{L_1}^{(i)}$ is the throat direction, while
${\bf e}_{\perp}^{(i)}$ denotes the corresponding perpendicular direction
in the orbital plane. A radial ray from the selected component is then
written as
\begin{eqnarray}
{\bf x}^{(i)}(r,\theta_{L_1})
&=&
{\bf x}_i
+
r\left[
\cos\theta_{L_1}\,{\bf e}_{L_1}^{(i)}
+
\sin\theta_{L_1}\,{\bf e}_{\perp}^{(i)}
\right].
\label{eq:radial_ray_L1}
\end{eqnarray}
With this convention, $\theta_{L_1}=0$ identifies the direction toward the
Roche-lobe throat at $L_1$, not the barycentric coordinate $x=0$. The
lateral regions of the lobe correspond approximately to
$|\theta_{L_1}|\simeq \pi/2$, while $|\theta_{L_1}|\simeq \pi$ describes
the side opposite to the throat.

For a given gravitational model, the radial boundary
$r_{\rm model}^{(i)}(\theta_{L_1})$ is defined as the first positive
solution of the critical-equipotential condition
\begin{eqnarray}
\Phi_{\rm eff}^{\rm model}
\left[
{\bf x}^{(i)}(r,\theta_{L_1})
\right]
&=&
\Phi_{L_1}^{\rm model},
\label{eq:radial_boundary_condition}
\end{eqnarray}
where
\begin{eqnarray}
\Phi_{L_1}^{\rm model}
&=&
\Phi_{\rm eff}^{\rm model}({\bf x}_{L_1}) ,
\label{eq:phi_L1_model}
\end{eqnarray}
and ${\bf x}_{L_1}$ is the model-dependent position of the inner Lagrange
point. The same construction is applied to the Newtonian reference case,
giving $r_{\rm Newt}^{(i)}(\theta_{L_1})$. The local fractional deformation
of the boundary is then measured by
\begin{eqnarray}
\Delta_{\rm rel}^{(i)}(\theta_{L_1})
&=&
\frac{
r_{\rm model}^{(i)}(\theta_{L_1})
-
r_{\rm Newt}^{(i)}(\theta_{L_1})
}{
r_{\rm Newt}^{(i)}(\theta_{L_1})
}.
\label{eq:delta_rel_theta}
\end{eqnarray}
This diagnostic compares the non-Newtonian and Newtonian Roche-lobe
boundaries along the same angular direction, measured from the same stellar
component, and therefore isolates the angular dependence of the
deformation.

The throat direction requires a separate treatment. At
$\theta_{L_1}=0$, the critical equipotential passes through the saddle
point $L_1$, where the two lobes meet. In a radial construction, rays
arbitrarily close to this cusp may become numerically ambiguous by
intersecting neighbouring branches of the same critical equipotential.
Therefore, in the residual plots we exclude a narrow angular interval
around $\theta_{L_1}=0$ from the connected curves and show the fractional
shift of the $L_1$ distance as an isolated marker.

Figure 5 displays the resulting residuals for the
Yukawa and power-law models. The main feature is that the largest
deformation does not necessarily occur at the throat. While
$\theta_{L_1}=0$ mainly tracks the displacement of $L_1$, the extrema of
$\Delta_{\rm rel}^{(i)}$ are typically found in the lateral sectors, around
$|\theta_{L_1}|\sim 80^\circ$--$110^\circ$. This indicates that the
modified potentials affect not only the position of the inner Lagrange
point, but also the transverse width and curvature of the critical
equipotential. For the parameter values shown, the power-law correction
produces larger residuals than the Yukawa model. The relative strength of
the deformation also depends on which lobe is considered: for small mass
ratios the effect is more pronounced around the lighter component, whereas
for larger mass ratios the dominant residual is shifted to the companion
lobe.

Figure 6 shows that the fitted Yukawa curve is essentially indistinguishable from the numerical Yukawa points across the mass-ratio interval explored, confirming that the multiplicative-polynomial ansatz captures the weak Yukawa deformation without spoiling the Newtonian scaling. The corresponding residuals, displayed in the top-right panel, stay within $\sim 1\%$ of unity and are comparable to the Newtonian deviations from Eggleton, indicating that the fit is accurate at the level of our numerical benchmark.

For the power-law model (here with $p=1$ and $\beta=0.05$), the deviations from Newtonian gravity are slightly larger but still smoothly varying in $q$. The $q$-power fit tracks the numerical points with a typical accuracy of a few per cent, while preserving the overall $R_L \propto q^{1/3}$ trend and the symmetry between $q$ and $1/q$. The residuals show that the largest corrections occur at the most extreme mass ratios, where the lobe is strongly asymmetric, but even there the fit remains well-behaved and monotonic.

On the other hand, a direct comparison with non-perturbative Newtonian deformations requires a stellar-structure model and is beyond the point-mass Roche framework adopted here. In
sufficiently close binaries, tidal, rotational, and finite-size effects may dominate over
the modified-potential corrections considered in this work. The present results should
therefore be interpreted as isolating the effect of changing the gravitational potential
itself. Observational relevance requires the fractional correction $\epsilon_Y$ or $\epsilon_P$ to be larger than, or at least comparable to, the uncertainty and deformation effects associated
with the conventional close-binary physics.

\begin{figure}[h!]
    \includegraphics[width=0.45\textwidth]{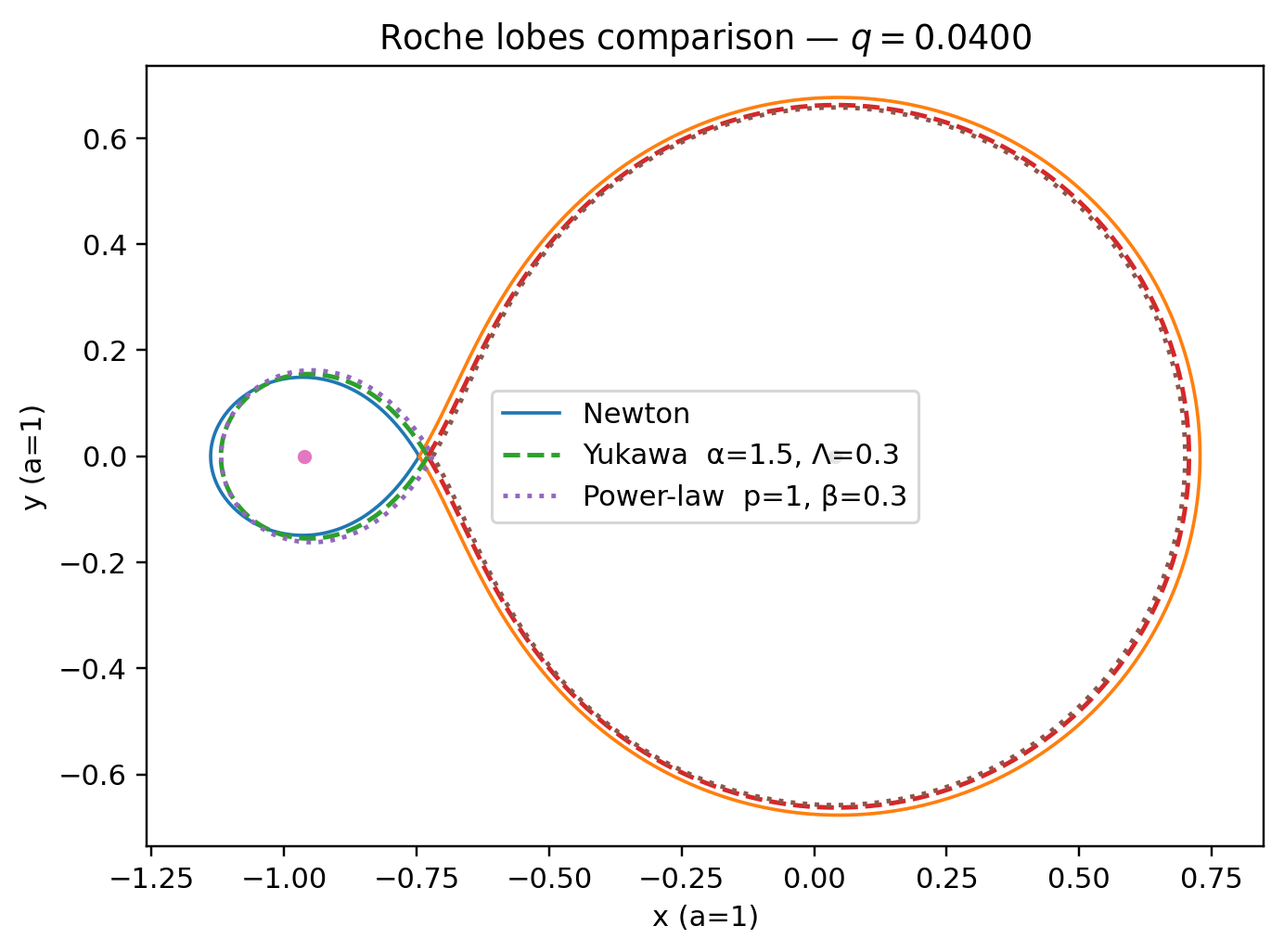}
    \includegraphics[width=0.45\textwidth]{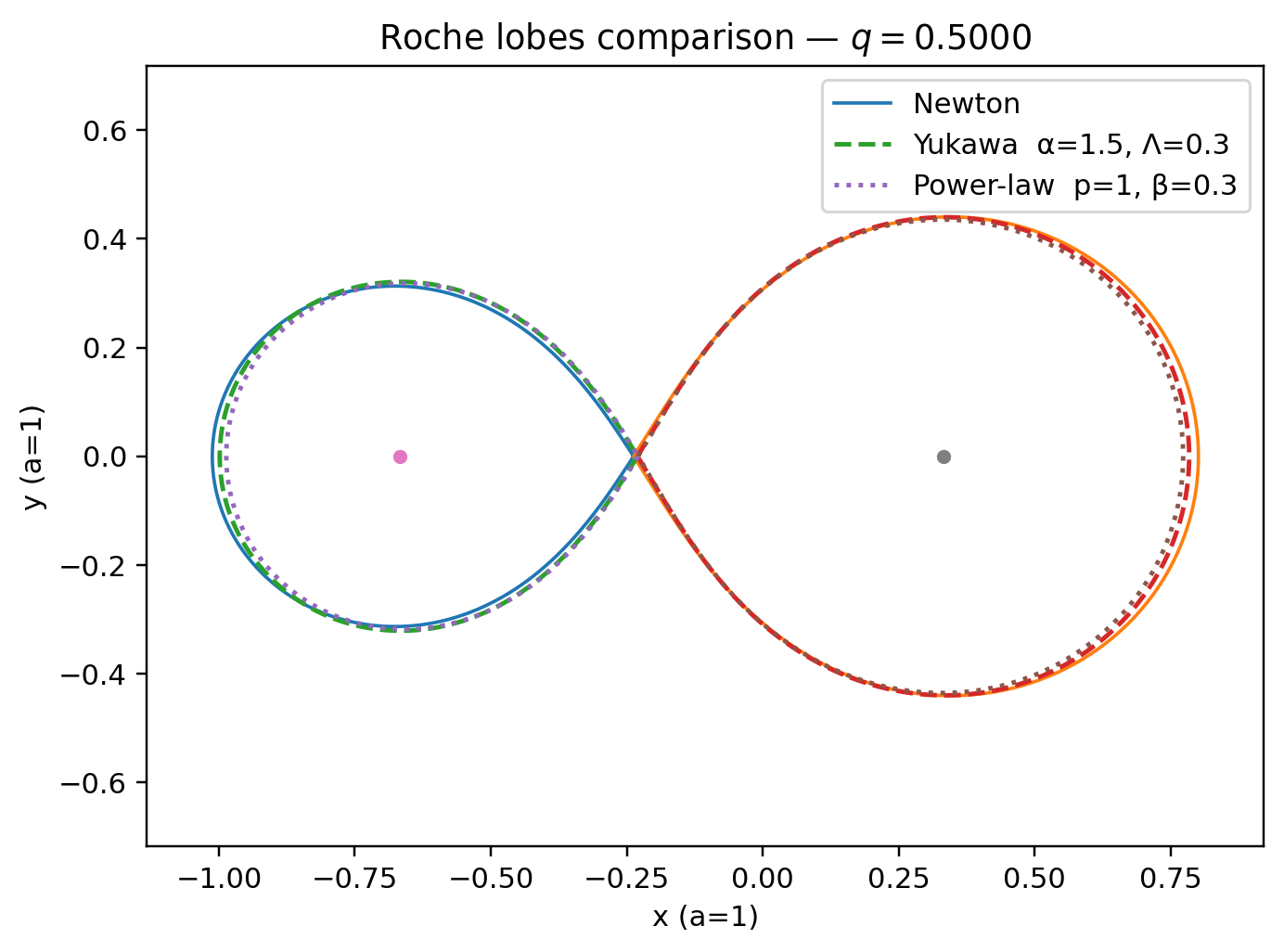}
    \includegraphics[width=0.45\textwidth]{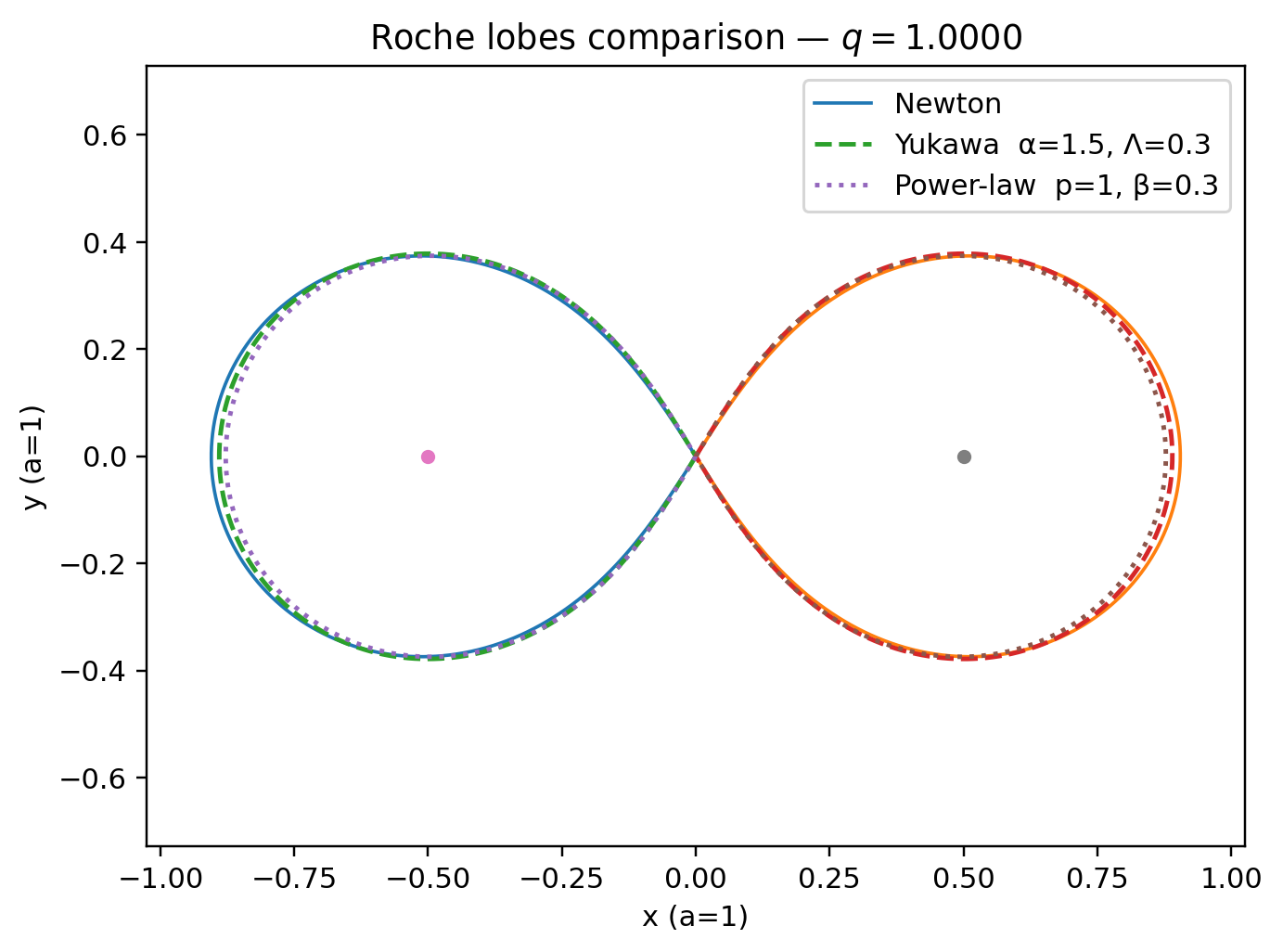}
    \includegraphics[width=0.45\textwidth]{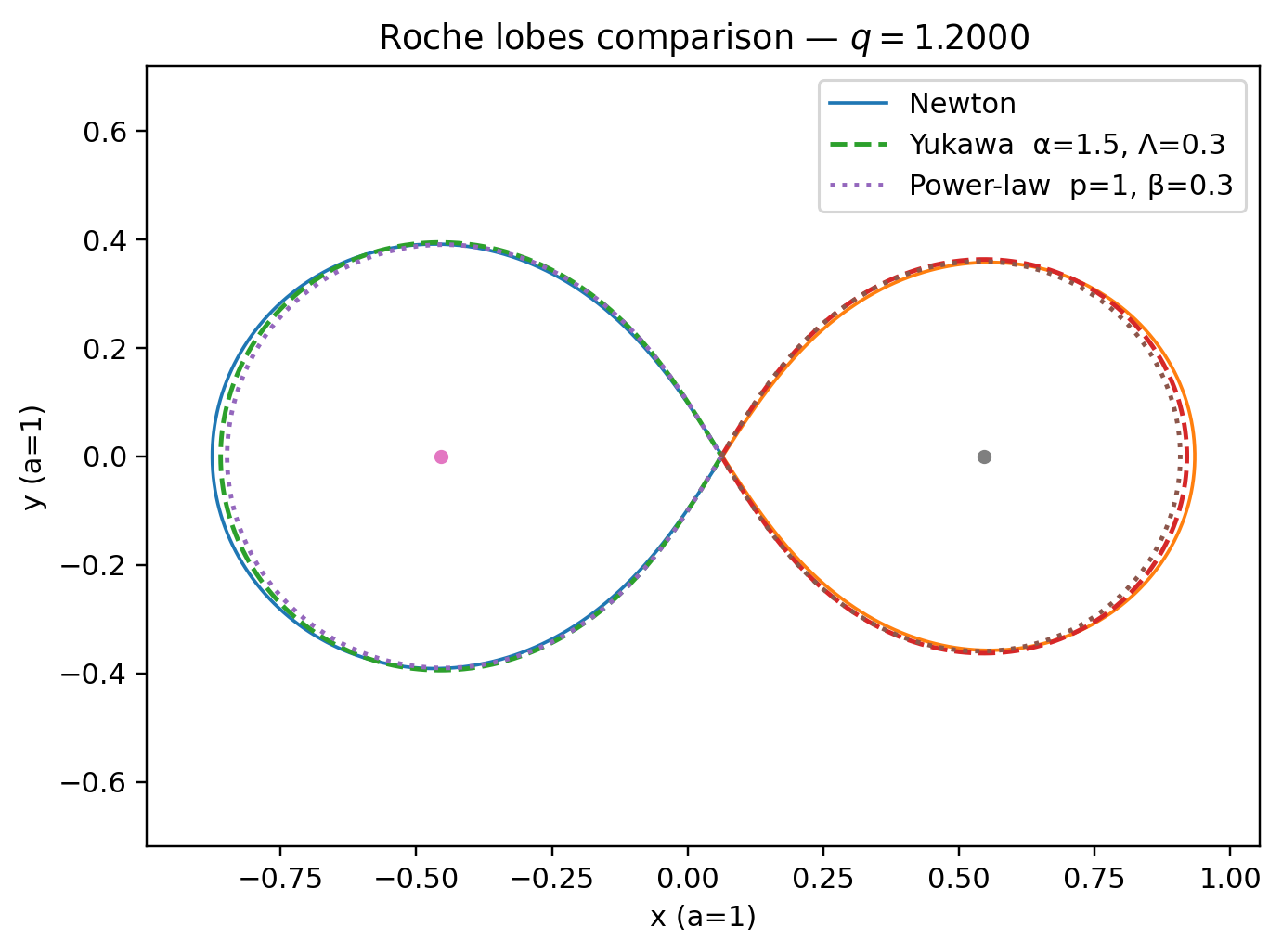}
    \caption{Two-dimensional Roche-lobe contours in the orbital plane for the Newtonian poten-
tial, the Yukawa potential, and the power-law potential, shown for representative
values of the mass ratio q. Solid, dashed, and dotted curves denote the Newtonian,
Yukawa, and power-law cases, respectively. The two colors distinguish the lobes
associated with the two binary components. The small markers indicate the point-
mass positions in barycentric coordinates. The angular frequency is computed from
the circular-orbit condition appropriate to each model.}
    \label{fig:rochea}
\end{figure}

\begin{figure}[h!]
    \includegraphics[width=0.45\textwidth]{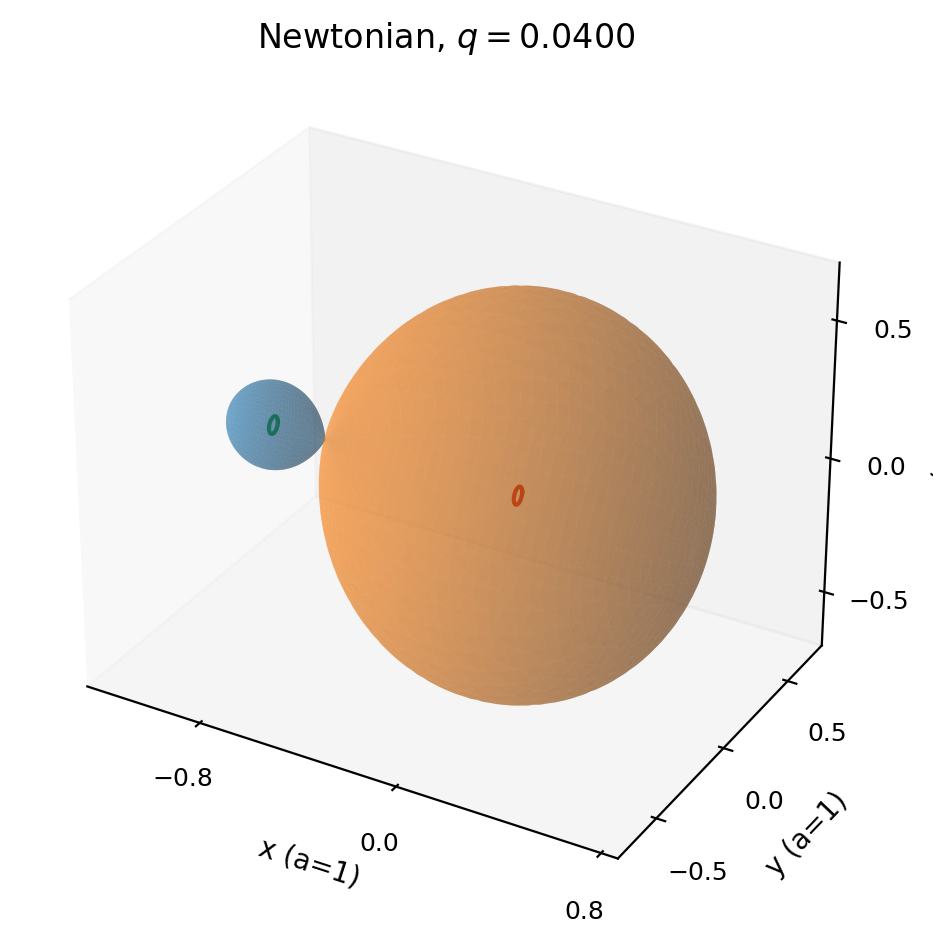}
    \includegraphics[width=0.45\textwidth]{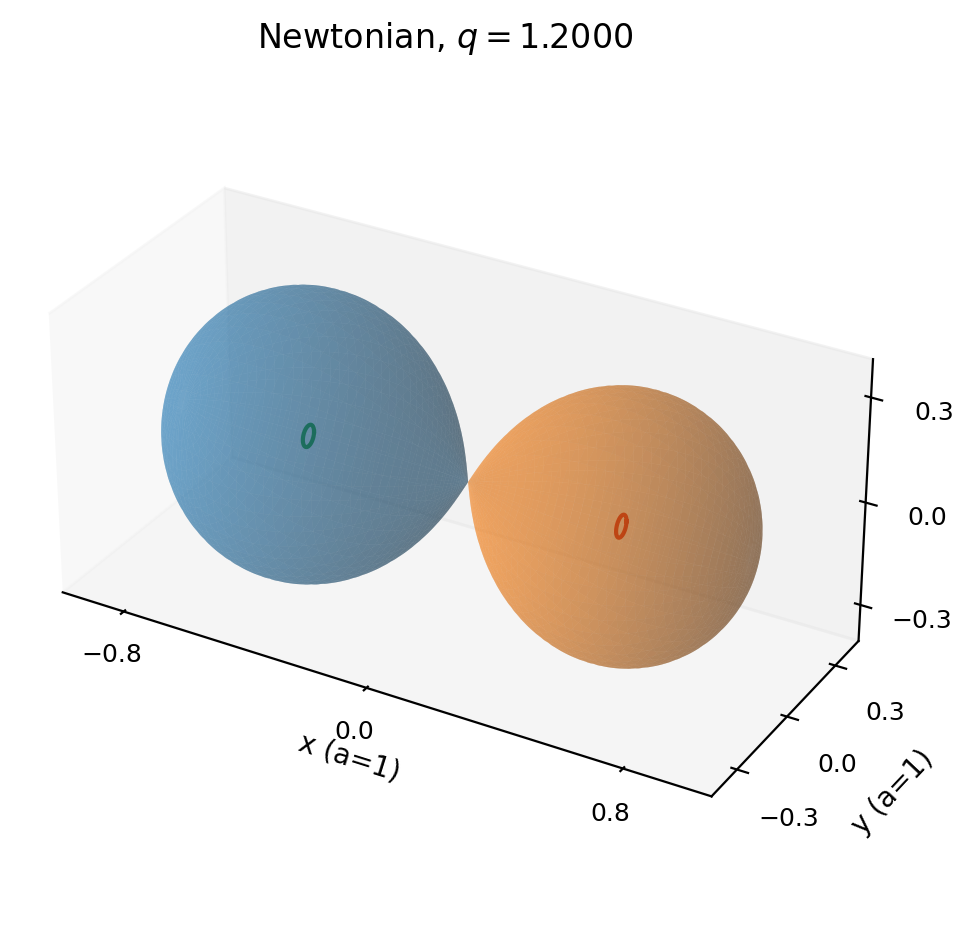}
    \caption{Three-dimensional reconstruction of the Newtonian Roche lobes for $q = 0.0400$ and
$q = 1.2000$. The surfaces are obtained from the critical equipotential passing through
L1 and define the corresponding volume-equivalent Roche lobes. The small rings
mark the barycentric positions of the two point masses and are shown only as visual
guides; they do not represent stellar radii.}
    \label{fig:rocheb}
\end{figure}

\begin{figure}[h!]
    \includegraphics[width=0.45\textwidth]{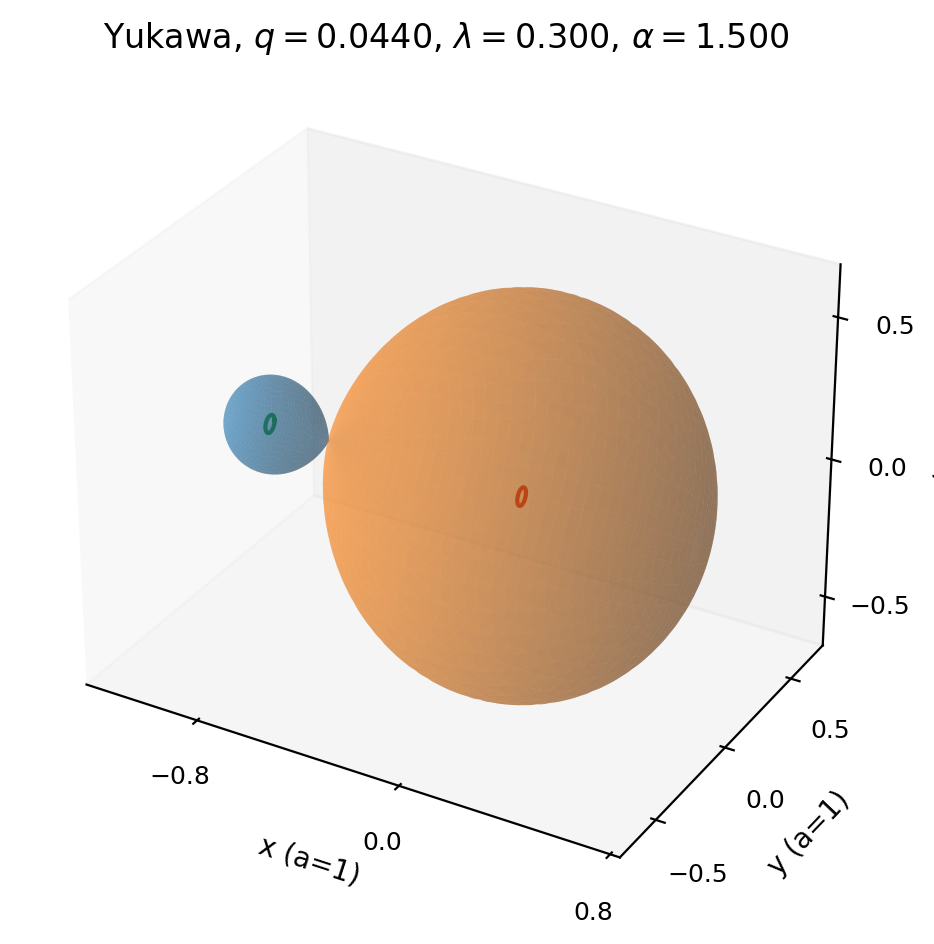}
    \includegraphics[width=0.45\textwidth]{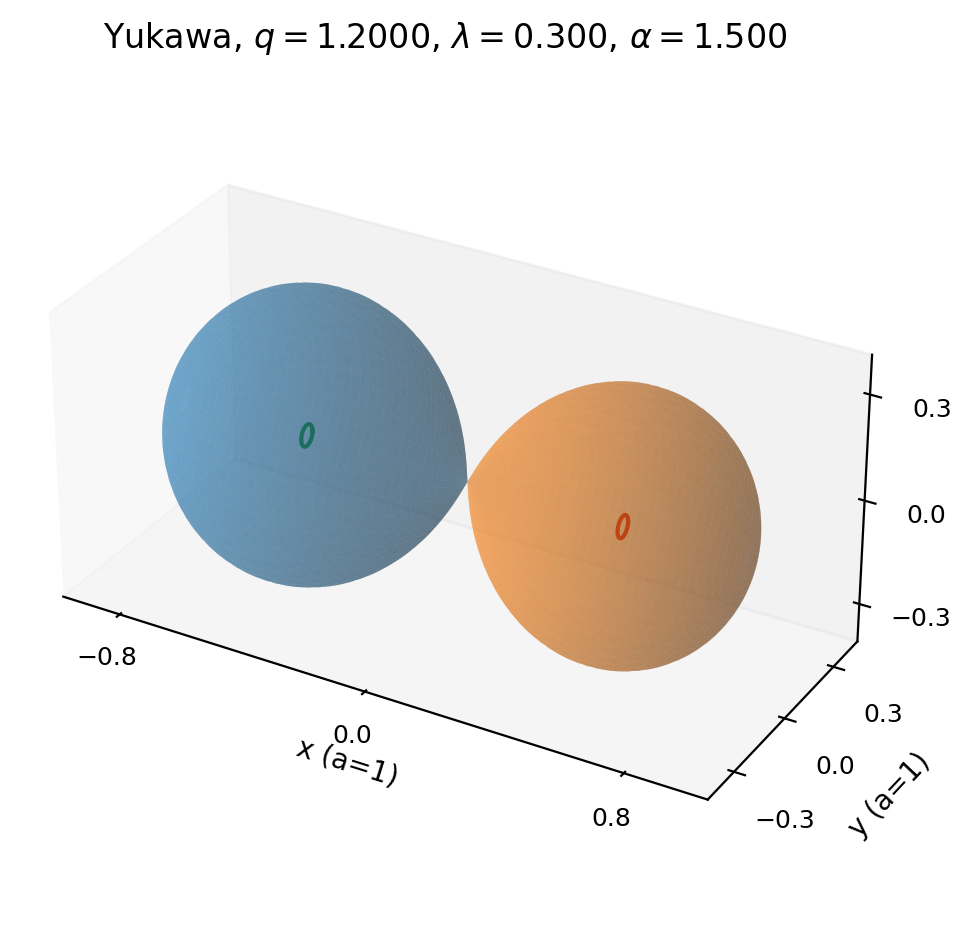}
    \caption{Three-dimensional reconstruction of the Roche lobes for the Yukawa potential, with
$ \Lambda = 0.3$ and $ \alpha = 1.5 $, for the same representative mass ratios as in the Newtonian
case. The small rings mark the barycentric positions of the two point masses and
are not physical stellar surfaces.}
    \label{fig:rochec}
\end{figure}

\begin{figure}[h!]
    \includegraphics[width=0.45\textwidth]{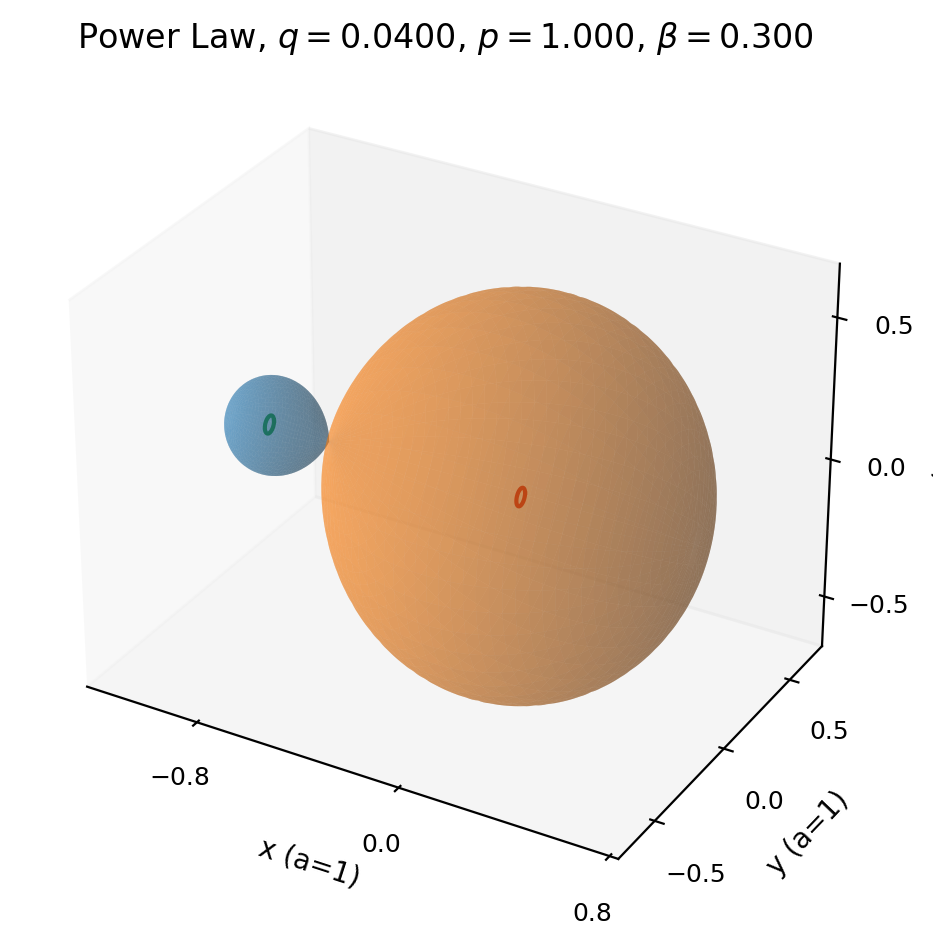}
    \includegraphics[width=0.45\textwidth]{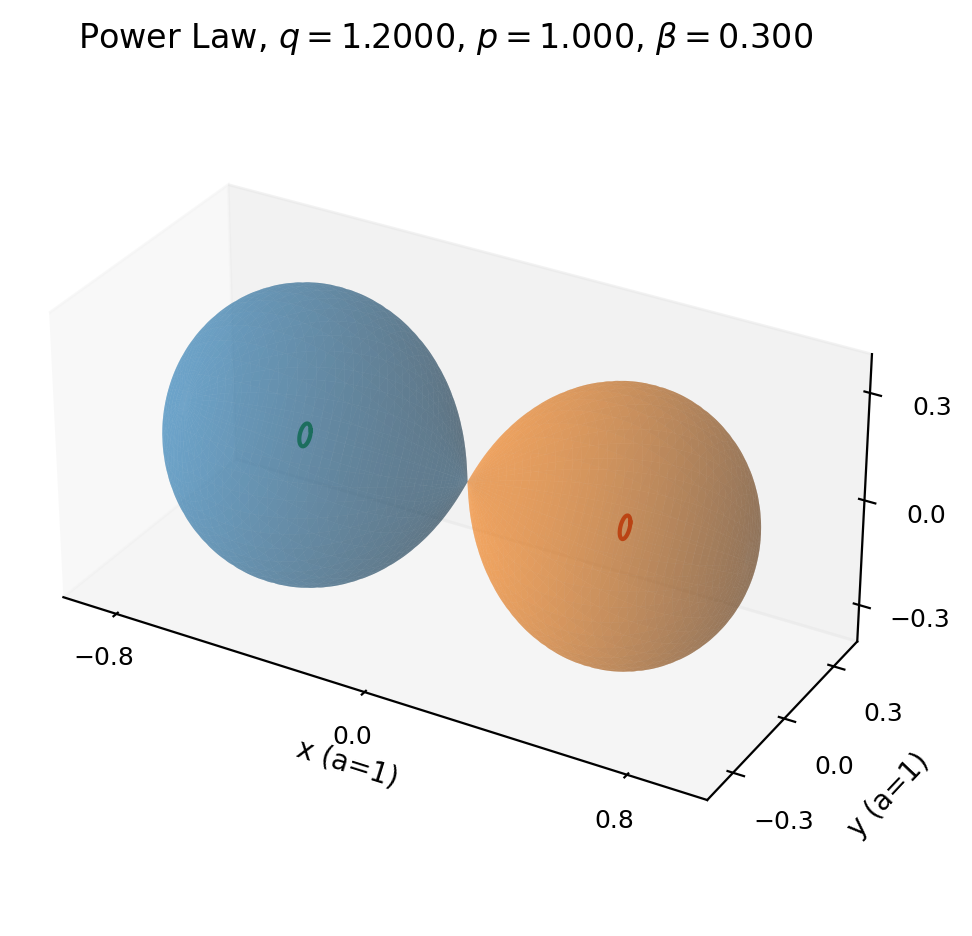}
    \caption{Three-dimensional reconstruction of the Roche lobes for the power-law correction to
Newtonian gravity, with $p = 1.0$ and $\beta = 0.3$. The small rings identify the positions
of the two point masses in the co-rotating barycentric frame and are included only
as visual markers.}
    \label{fig:roched}
\end{figure}

\begin{figure}[h!]
    \includegraphics[width=1.000\textwidth]{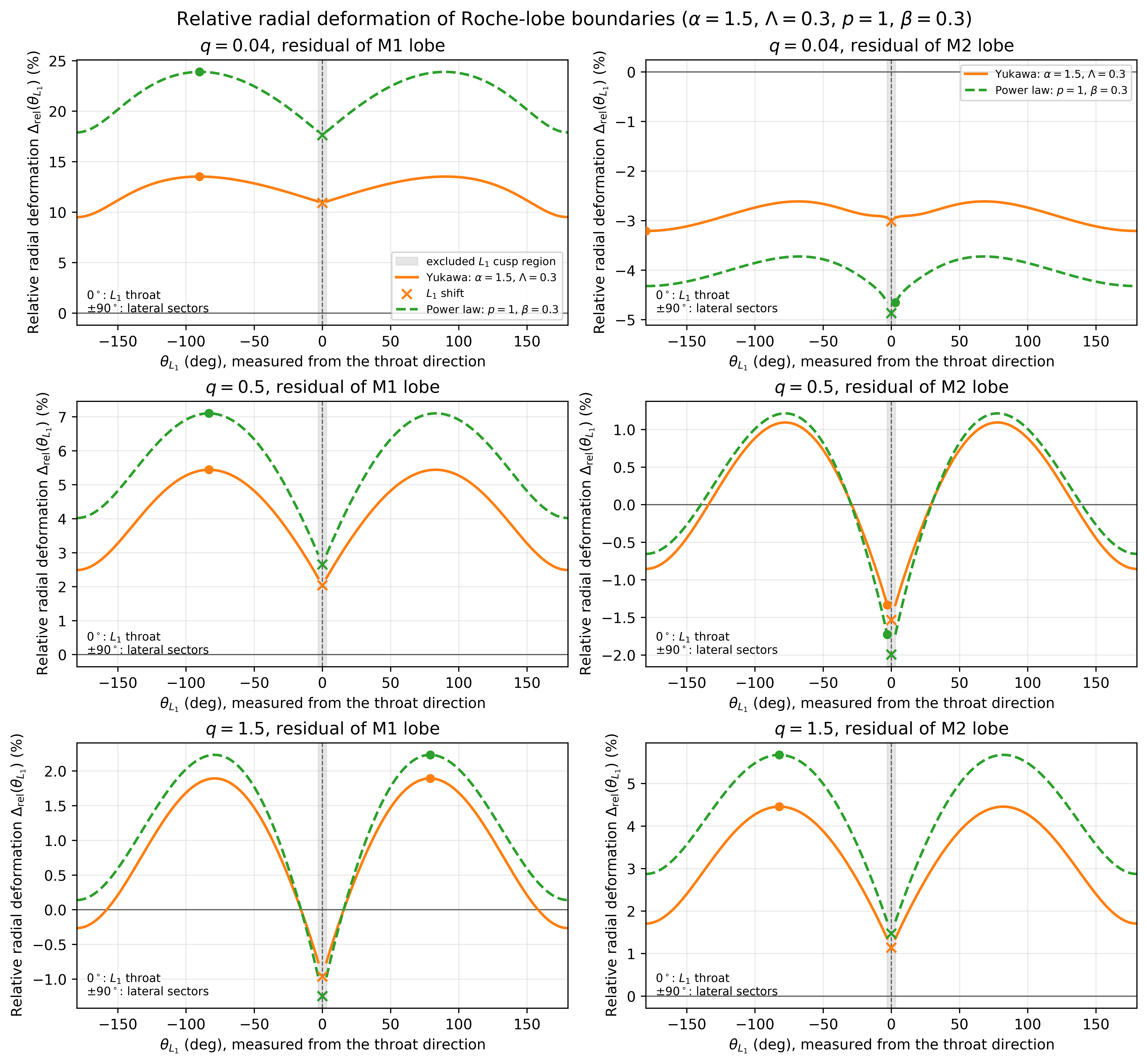}
    \caption{
Relative radial deformation of the non-Newtonian Roche-lobe boundaries
with respect to the Newtonian case in the orbital plane. The left and
right columns show the residuals for the lobes around $M_1$ and $M_2$,
respectively, for different mass ratios $q$. The plotted quantity is
$\Delta_{\rm rel}(\theta_{L_1})$. The angle
$\theta_{L_1}$ is measured from the direction pointing from the selected
component to the Roche-lobe throat at $L_1$. The grey region around
$\theta_{L_1}=0$ marks the excluded cusp region, while the crosses indicate
the fractional shift of the $L_1$ distance. Circular markers denote the
maximum absolute residual along the smooth part of the boundary. The
power-law correction produces larger distortions than the Yukawa model for
the parameter values shown, and the largest deviations typically occur in
the lateral sectors of the lobes rather than exactly at the throat.
}
    \label{fig:roched}
\end{figure}

\begin{figure}[h!]
    \includegraphics[width=0.45\textwidth]{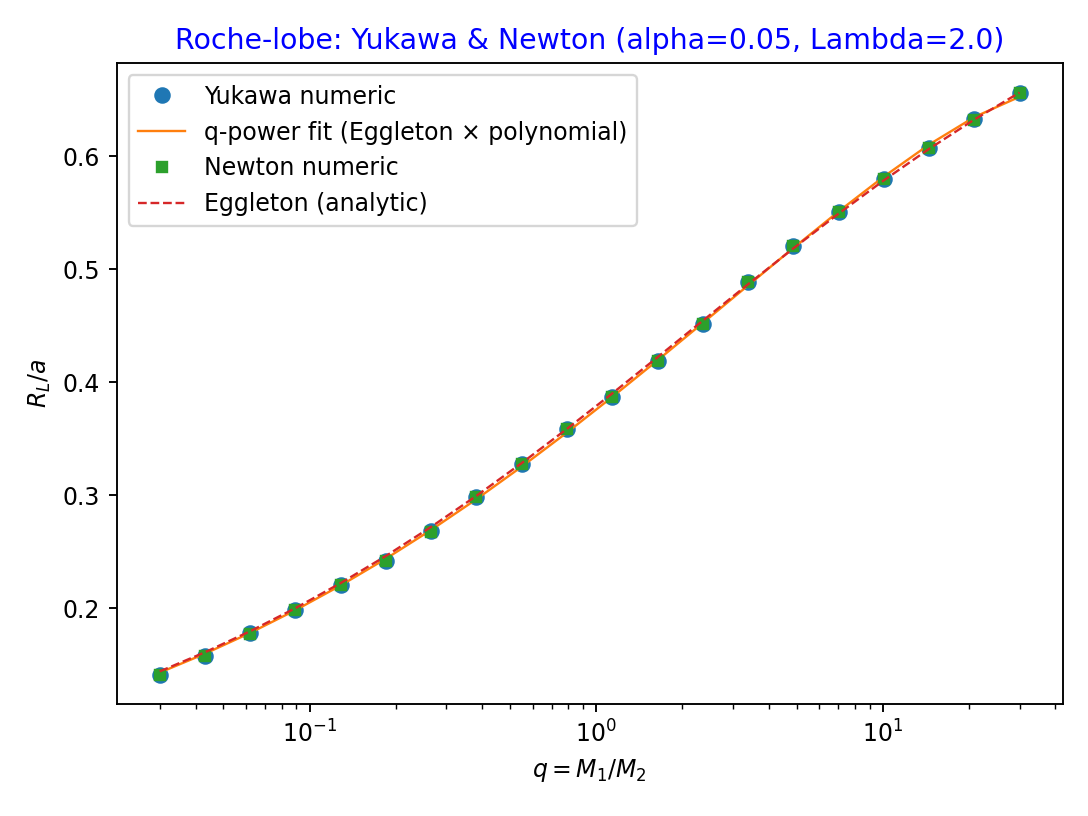}
    \includegraphics[width=0.45\textwidth]{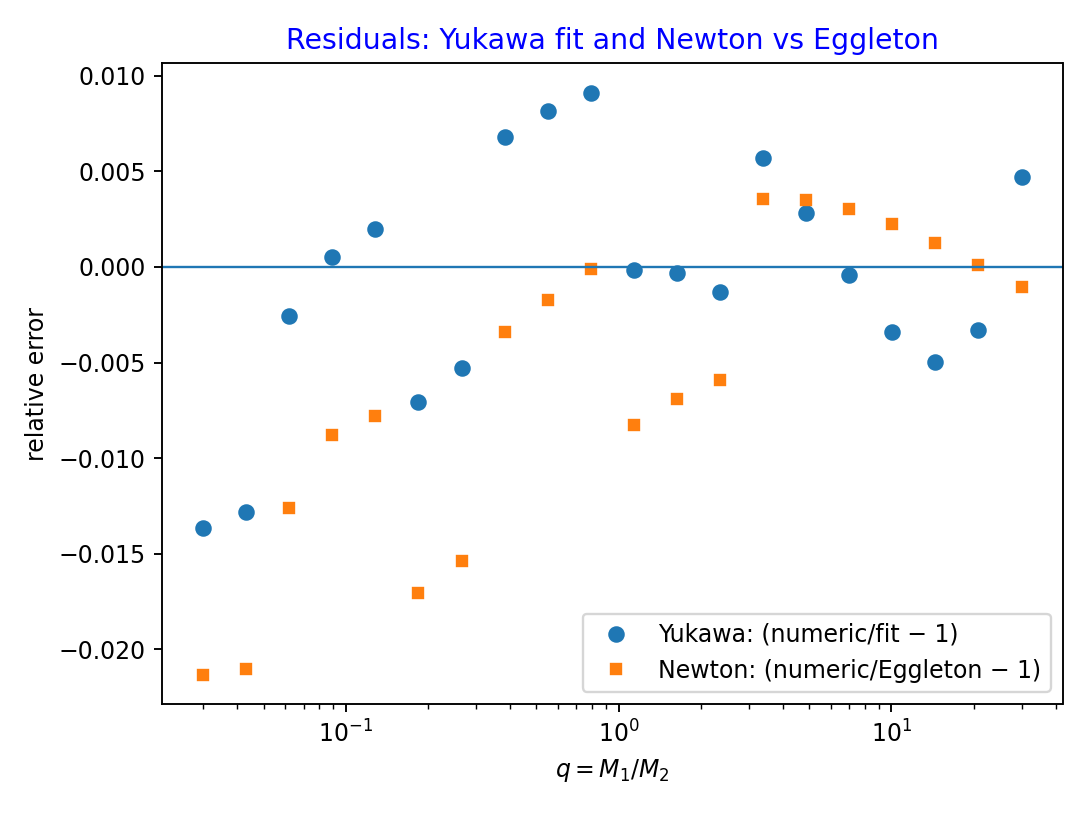}
    \includegraphics[width=0.45\textwidth]{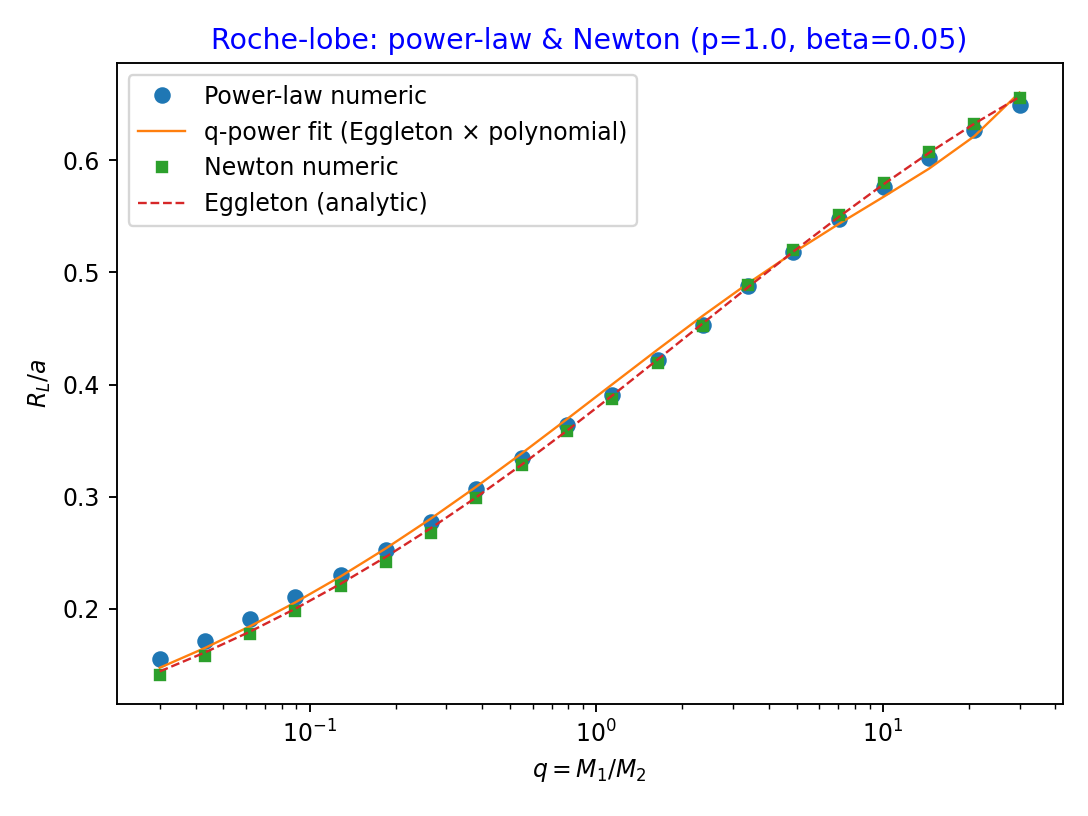}
    \includegraphics[width=0.45\textwidth]{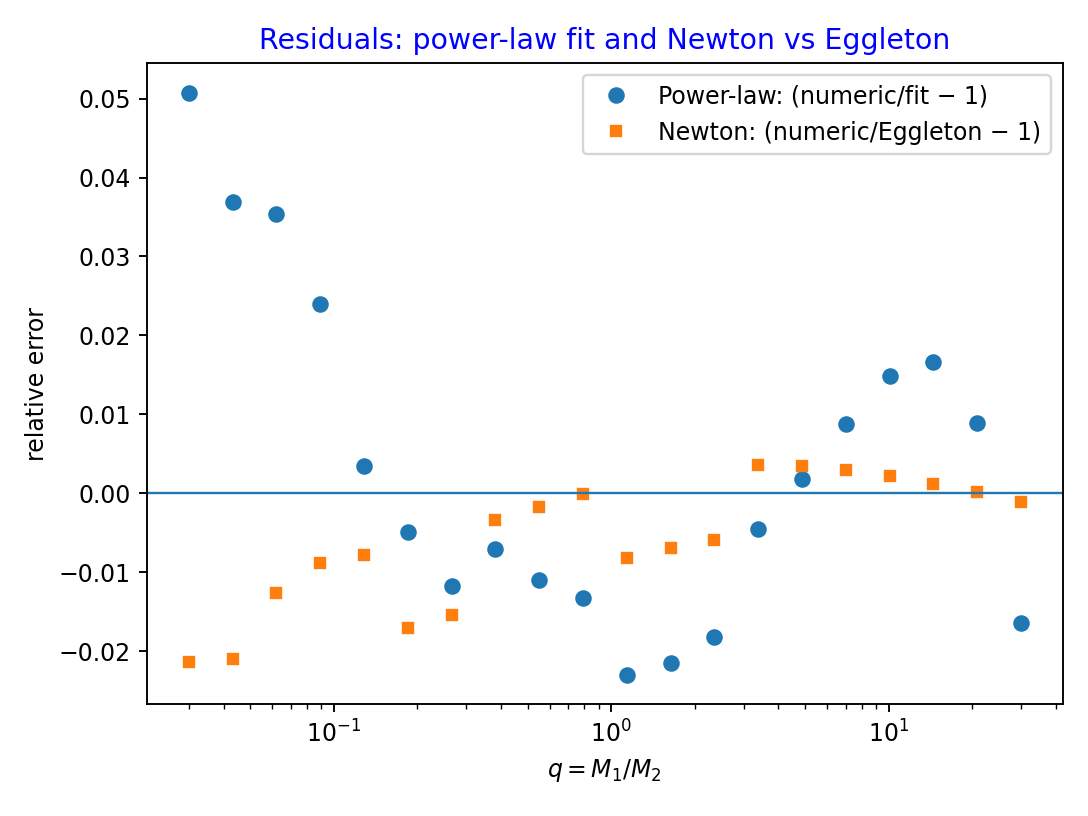}
    \caption{Eggleton-like fitting formulas for the equivalent-volume Roche-lobe radius in the
Yukawa and power-law models. Left panels show the numerical values of $R_L/a$
together with the fitted curves, while right panels show the relative residuals. The
fits preserve the Newtonian small-q scaling and provide a compact approximation to
the numerically computed non-Newtonian Roche-lobe radii.}
    \label{fig:roche}
\end{figure}

\section{Conclusions}

In this work, we have revisited the classical construction of the Roche lobe in Newtonian gravity and extended it to two families of non-Newtonian potentials of broad phenomenological interest, namely Yukawa-type and power-law corrections. Starting from the $L_1$ stationarity condition and its local expansion, we derived the small–$q$ asymptotics and then used a tangent-field plus Newton-projection scheme to compute the full two-dimensional equipotential contours and the corresponding three-dimensional Roche-lobe volumes in a transparent and numerically robust way.

For the Newtonian case, we recovered the standard behavior of the equivalent-volume radius and used Eggleton’s formula as a benchmark. We then showed that the same numerical machinery can be applied almost unchanged to the Yukawa and power-law potentials, with the only modification being the effective potential and its gradients. The resulting two-dimensional contours, displayed for several mass ratios, confirm that there exist regions in parameter space where the Yukawa and power-law lobes are nearly indistinguishable from the Newtonian Roche lobe, as well as regimes where the deviations become significant. The corresponding three-dimensional reconstructions illustrate how the overall shape and volume of the lobe respond smoothly to variations in the Yukawa parameters $(\alpha,\lambda)$ and in the power-law parameters $(p,\beta)$.

On top of these numerical results, we constructed Eggleton-like fitting formulas for the equivalent-volume Roche-lobe radius in the non-Newtonian models. For the Yukawa case, we identified the relevant cubic scale from the small–$q$ expansion, which involves the near-body factor $\Upsilon_0$, the orbital factor $\Upsilon_1$ entering $\Omega^2$, and a small gradient correction $\zeta$. We then built a multiplicative ansatz on the Eggleton baseline, expanded in fractional powers of $q$, with coefficients $d_i(\lambda,\alpha)$ calibrated by least squares against the numerical $R_L/a$ over a grid in $(\alpha,\lambda)$. An analogous strategy was implemented for the power-law potential, with coefficients $d_i(p,\beta)$ that reduce continuously to the Newtonian limit as $\beta \to 0$. In both cases, the residuals remain small and well behaved across the range of mass ratios explored, indicating that the fits capture the main dependence on $q$ and on the non-Newtonian parameters without introducing spurious features.

These results show that both Yukawa-modified gravity and power-law
corrections can be incorporated into Roche-lobe calculations in a way that
is simultaneously numerically transparent and computationally efficient.
The explicit 2D and 3D constructions provide a direct visualization of how
alternative gravities reshape the critical equipotential surfaces. In
addition, the angular residual tool introduced in Figure 5 makes these
deviations quantitatively visible even when the Roche-lobe surfaces appear
nearly degenerate. The residual curves show that the largest deformations
do not necessarily occur at the throat itself, but often in the lateral
sectors of the lobe, reflecting changes in the transverse curvature and
width of the critical equipotential. For the representative parameters
considered here, the power-law correction produces larger angular
distortions than the Yukawa model. The calibrated fitting formulas then
supply fast approximations suitable for stellar-evolution and
population-synthesis applications. Although the examples presented here
focus on modest deviations from Newtonian gravity, the methodology can be
straightforwardly extended to broader regions of parameter space and to
other parameterizations of non-Newtonian forces.

The present results should be viewed as a controlled phenomenological extension of the classical Roche construction rather than as a complete model of close-binary structure. The fitting formulae derived here provide fast approximations to the Roche-lobe radius once a modified potential and a parameter range have been specified. Their direct use in binary-evolution or population-synthesis calculations will require additional physical input: in particular, a motivated range for the non-Newtonian length scale or power-law amplitude, and a comparison with finite-size, tidal, rotational, and non-perturbative Newtonian effects. Within these limitations, the formalism developed here offers a useful way to isolate how changes in the gravitational potential alone deform the critical Roche equipotential and its volume-equivalent radius.

\section*{Acknowledgments} 

I would like to thank Maria Margarita and Miguel Amado for their continuous inspiration and support, and the Foundation for Research Support of Espírito Santo (FAPES) for the partial support for the present work.

\end{document}